# Imaging the effect of drive on the low-field vortex melting phenomenon in $Ba_{0.6}K_{0.4}Fe_2As_2$ single crystal

Ankit Kumar[1], Amit Jash[1], Tsuyoshi Tamegai[2], S. S. Banerjee[1,*]

[1]Department of Physics, Indian Institute of Technology, Kanpur-208016, India

[2]Department of Applied Physics, The University of Tokyo, Hongo, Bunkyo-ku, Tokyo 113-8656, Japan

[*]Email: satyajit@iitk.ac.in

**Abstract**

Self-field imaging of current distribution in $Ba_{0.6}K_{0.4}Fe_2As_2$ superconductor is used to study the effect of drive on the low-field vortex solid to liquid melting phase transformation. At low fields, the current induced drive on the vortices aids in thermally destabilizing the vortex state thereby shifting the low field melting phase boundary. We show that the current induced drive shifts solid–liquid boundaries and prepones the vortex melting phenomenon compared to the equilibrium situation. The analysis shows that for currents above 50 mA, Joule heating effects shift the melting line while below 50 mA, an effective temperature concept for driven system viz., a drive dependent shaking temperature, explains the shift. The observation of a transformation from inhomogeneous to homogeneous current flow in the sample at low fields as a function of the driving force is reconciled via an inverse dependence of the shaking temperature on vortex velocity, which is incorporated in our analysis.

Interacting particles driven across a random pinning media result in different non-equilibrium (NEQ) phases [1,2]. Some established general NEQ phases are the fluctuating and quiescent phases [1,3,4] for driven matter. Study of systems organizing themselves randomly into these distinct NEQ phases as a function of drive is important, as they aid in developing the general framework for understanding NEQ phase transitions. The moving vortex state in a type-II superconductor is a convenient prototype of driven NEQ system. It is akin to an elastic manifold driven though a random pinning environment [5,6,7]. In the presence of an applied field $B_a$, the current $I$ induced Lorentz force ($F_L \propto IB_a$) drives the vortices in a superconductor. In realistic superconductors the vortices are pinned and they depin at a critical threshold current value $I_c$ when $F_L$ exceeds the pinning force. At different drives the vortex flow can organize into different moving phases, like, the elastic, plastic moving Bragg glass phase, and the driven jammed phase [5,6,7,8,9,10,11,12,13,14,15]. It may be mentioned here that most of these studies have been done (or performed) at high $B_a$ where vortices are strongly interacting, viz., intervortex spacing $a_0 \propto \sqrt{\frac{\phi_0}{B_a}} < \lambda$, $\phi_0$ is the flux quantum, and $\lambda$ is the penetration depth. Studies on driven vortex matter show that the pinning excites topological defects [16] in the moving phase, leading to a moving glassy vortex phase [17]. At larger driving forces the moving state is a well-ordered moving phase [5-8] where vortices maintain approximately fixed correlation with their neighbours. A model to understand how pinning affects the moving vortices states that [18,19], for moderate vortex velocity $v$, pins cause a fluctuation in the velocity of the vortices as they move past the pinning centres. Thus, when a collection of vortices is driven across a random pinning environment, they will intrinsically possess $v$ fluctuations. In the frame of reference of the moving vortices, the fluctuating NEQ state can be described with the concept of an effective temperature called the 'shaking temperature' ($T_{sh}$) [18,19]. The $T_{sh}$ is a NEQ concept and is different from the equilibrium temperature. The $T_{sh}$ has an inverse $v$ dependence and depending on how large or small $T_{sh}(v)$ is, determines the degree of order present in the moving vortex phase. While the concept of $T_{sh}(v)$ is useful, its quantitative experimental validation is absent. However, the concept of $T_{sh}$ has been applied to qualitatively explain features of plastic and elastic flow found in electrical transport (*I-V*) measurements performed on the vortex state [6,8-11]. To the best of our knowledge, the effect of $T_{sh}$ on phase transitions in the vortex matter under low-drive conditions, have not been well explored experimentally.

It may be mentioned here that in a static vortex lattice, loss of spatial correlations in an ordered vortex lattice precipitated either by, enhanced thermal fluctuations or enhanced pinning effects, leads to the formation of a vortex liquid [20,21] or vortex glass phases, respectively [5,6]. Theoretically, thermal fluctuations produce vortex melting at both high and low fields [20,21]. This occurs due to a weakening of the elastic moduli of the vortex lattice near the upper critical field ($a_0 < \lambda$), as well as in the weakly interacting regime near the lower critical field, where $a_0 > \lambda$. Recent experiments have found evidence for a low-field vortex liquid and a glassy phase in the static vortex matter of $Ba_{0.6}K_{0.4}Fe_2As_2$ single crystals [Ref. [22]]. Here using the self-field differential magneto-optical imaging (MOI$_{SF}$) technique, we image current distribution at low applied fields ($B_a$) in the same $Ba_{0.6}K_{0.4}Fe_2As_2$ single crystal of Ref. [22]. We see that at low $B_a$ the current distribution across the sample is non-uniform. The inhomogeneous flow arises because of the circulation of currents preferentially around a dilute vortex liquid-solid boundary formed as the vortex state melts with lower current in the liquid. With varying $I$, we see a shift of the low field melting phase boundary to lower temperatures and fields. We show that in the presence of $I$, vortex melting can be observed at very low fields, which was not possible when $I = 0$. This observation suggests that currents aid in thermally destabilizing the low-field glassy phase into a vortex liquid phase [Ref. [22]]. Our analysis shows that the drive-induced lowering of the vortex melting temperature occurs due to two effects. For $I \geq 50$ mA, Joule heating effect leads to a lowering of the dilute vortex melting temperature $T_m(I)$ w.r.t. $T_m(I = 0)$. For $I < 50$ mA, the lowering of melting temperature is explained by considering the effects of shaking temperature ($T_{sh}$). With varying drive we observe a transformation from inhomogeneous current distribution in the sample into a state with uniform current distribution. We believe this transformation suggests the inverse dependence of $T_{sh}$ magnitude on vortex velocity.

For our study, we use a high-quality optimally doped single crystal of $Ba_{0.6}K_{0.4}Fe_2As_2$ with dimensions $1.7 \times 1.2 \times 0.025$ mm$^3$ and $T_c = 38$ K [22,23]. In all our experiments, the applied magnetic field ($B_a$) was parallel to the crystallographic $c$ axis. Note that all measurements are for a field-cooled sample. The schematic in Fig. 1(a) shows Cr(10 nm)/Au(100 nm) current pads (yellow) with contact resistance ~10 m$\Omega$, DC-sputtered on the sample (blue). The flow of electrical current in $Ba_{0.6}K_{0.4}Fe_2As_2$ is imaged via the self-field generated by the current using the self-field magneto-optical imaging technique (MOI$_{SF}$) [24,25]. Briefly, in MOI$_{SF}$ the difference of magneto-optical images captured with $I$ sent into the sample in two opposite directions is captured $N$ times, and the difference is averaged $M$ times, MOI$_{SF}$ =

$\frac{1}{M}\sum_{j=1}^{M}[\sum_{i=1}^{N}\frac{[MOI_i(I+)-MOI_i(I-)]}{N}]_j$, (usually, we choose $N = M = 10$). The MOI$_{SF}$ images [Fig. 1(a)] is a map of the self-field, $B_z(x, y)$ generated by the current, where the $(x, y)$ represents the pixel co-ordinate in a magneto-optical image. A numerical inversion routine [24,26] is used to convert the self-field $B_z(x, y)$ map into a current density $j(x, y)$ map.

Figure 1(a) main panel shows a MOI$_{SF}$ image taken at $T = 29$ K, $B = 7$ G and $I = 40$ mA. The magneto-optical contrast at the edges of the contact pads outside the sample, as well as at the edges of the sample, are bright and dark, which corresponds to self-field $B_z$ pointing out of or into the plane of the figure, respectively. Over the sample surface, the contrast is uniformly grey.

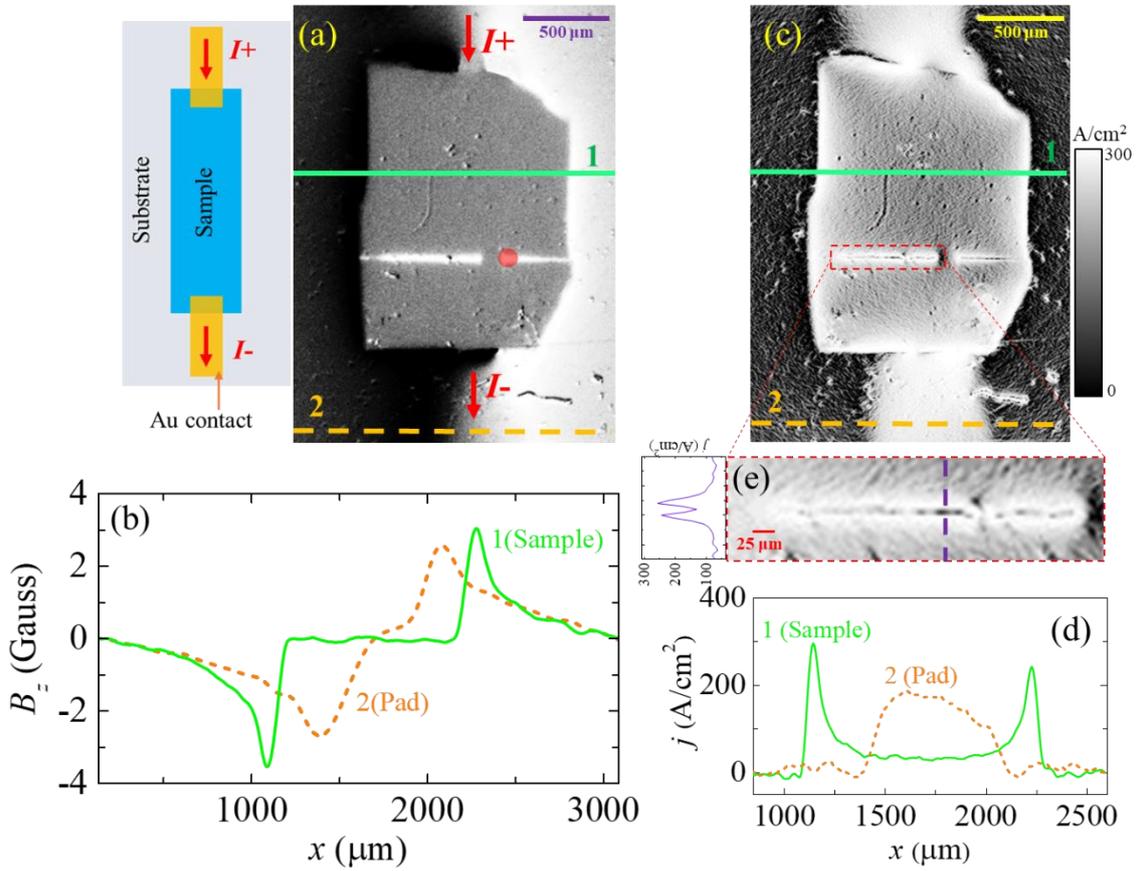

FIG. 1. (a) Left figure schematically shows current pads on the sample. Main panel (a) shows the MOI$_{SF}$ image taken at $T = 29$ K, $B = 7$ G and $I = 40$ mA (see brown arrow). (b) Self-field $B_z$ measured across the line, over the sample (green color), and current pad (dashed orange color) in Fig. (a). (c) Figure shows $j(x, y)$ of Fig. (a). (d) $j(x)$ measured across the line drawn over the sample (green color), and the current pad (dashed orange color). (e) Expanded view of the red dashed rectangular region. The graph shows $j(x)$ modulation along the purple dashed line.

Figure 1(b) shows the $B_z(x)$ profile measured along the dashed (orange) line and solid (green) line drawn across the MOI$_{SF}$ image [see Fig. 1(a)] over the current pad and sample surface, respectively. The maxima's in $B_z(x)$ profile in Fig. 1(b) are found at the edges of the sample and the current pad. Figure 1(c) shows $j(x, y)$ image obtained by inverting [24,26] the $B_z(x, y)$ image of Fig. 1(a). The orange dashed curve in Fig. 1(d) shows a nearly uniform current sheath flowing across the width of the current pads (curve marked 2). Thus, the nature of the $B_z$ profile in Fig. 1(b) measured across the current pad corresponds to a uniform current sheath flowing in the pad. In Fig. 1(d), the $j(x)$ profile across the sample (green curve), shows relatively strong currents flowing along the sample edges (~200–250 A.cm$^{-2}$) and relatively smaller amount of $j$ (~50 A.cm$^{-2}$) in the sample bulk, due to strong geometric barrier effects [22,27,28,29]. In Fig. 1(a) we see region with bright (white) contrast in the sample. This is the same region seen in Fig. 3 of Ref. [22], where low-field vortex melting occurs (also in Supplementary Material Fig. 2, shows bright contrast in the images correspond to enhancement in local field vortex density due to low field vortex melting with $I = 0$ mA in this region). Figure 1(e) is an expanded view of a portion inside the dashed rectangular region of Fig. 1(c), showing a region with bright contrast in $j(x, y)$ on its periphery with a darker contrast inside it. The profile $j(x)$ is measured along the (purple) dashed vertical line in Fig. 1(e) and shows high current density present around the melted region with a lower current density present in the central darker region. Such a feature is expected for a vortex liquid puddle surrounded by a solid phase, as there will be a relatively higher local resistance inside the liquid phase compared to the periphery with the liquid-solid interface. Hence, currents will flow along the solid-liquid interface with lesser current flowing in the liquid interior. This leads to the dip in $j(x)$, as seen in Fig. 1(e).

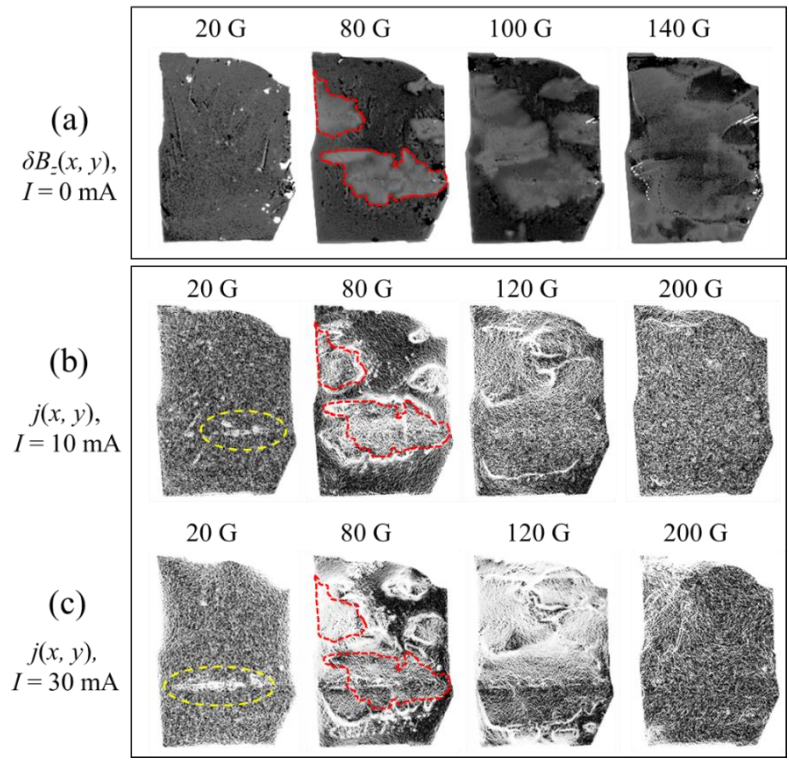

FIG. 2. (a) $\delta B_z(x, y)$ images at 17.2 K for different $B_a$. (b-c) $j(x, y)$ images for $I = 10$ and 30 mA sent into the sample at 17.2 K for different $B_a$. The red dashed contours compare changes in the liquid-solid contours with increasing $I$.

The first row of images in Fig. 2(a) is a map of changes in local field, i.e., $\delta B_z(x, y)$ produced in response to 1 G modulation in $B_a$ [see Ref. [22]] with $I = 0$ mA. At 17.2 K with $I = 0$ in Fig. 2(a), the bright regions in the images are associated with a local enhancement ($\delta B_z$), i.e., jump in local vortex density [22] produced due to vortex melting. We see the nucleated melted vortex liquid regions spread out across the sample with increasing $B_a$. Next, we send in $I$ and compare the images at these $B_a$ to study how the vortex melting gets affected. Instead of using self-field images, it is instructive to see Figs. 2(b) and (c) the $j(x, y)$ images at 17.2 K for $I = 10$ and 30 mA, respectively. Note the $j(x, y)$ image of Fig. 2(b) clearly shows just the beginning of bright melting patches at 20 G for 10 mA, which becomes more prominent at 30 mA (see within the yellow dashed curve). Note from Fig. 2(a) that at $I = 0$ mA there is no evidence of melting starting at 20 G. Thus, with the application of current, the melting seems to begin from a lower $B_a$. For $B_a = 40$ G and above, the bright contours in $j(x, y)$ images are currents flowing along the liquid-solid boundaries. We observe that as $I$ is increased (keeping $B_a$ fixed) [for example compare 10 mA, 80 G [Fig. 2(b)] with 30 mA, 80 G [Fig. 2(c)] image], the vortex liquid-solid boundary is seen to expand with $I$ w.r.t the liquid-solid contours in Fig. 2(a) at 80 G with $I = 0$ mA [marked as the red dashed curve in Figs. 2(a), (b), and (c)]. We note from Figs. 2(b) and (c) that from 120 Oe onwards, the current distribution in the sample begins to get uniform. Thus, we see that the onset of melting is shifted with the application of current. The preponing of melting of the vortex state with current, leads to the observed expansion of the vortex solid-liquid contours with $I$ in presence of $B_a$.

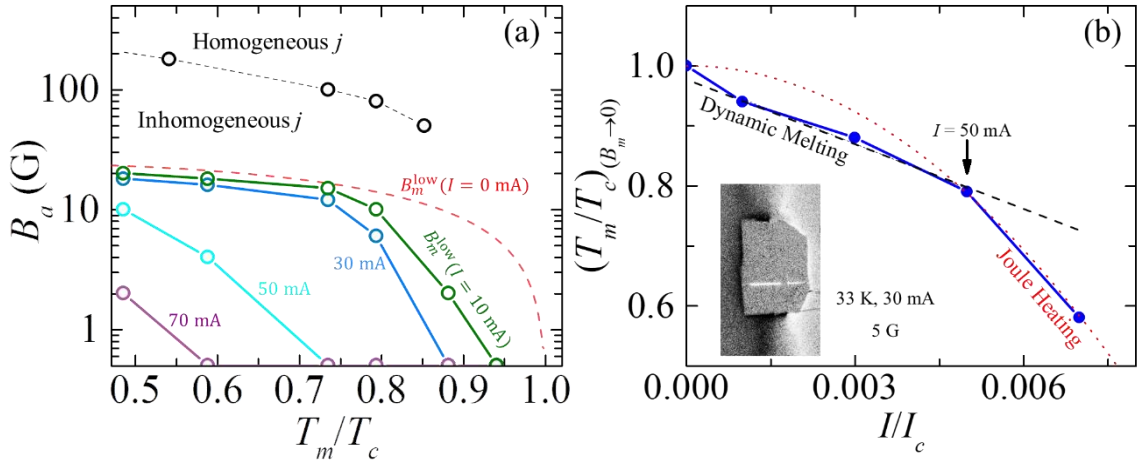

FIG. 3. (a) Figure shows the melting field $B_m$ and temperature $T_m$ where low field melting occurs at the point marked as a red • in Fig. 1(a), in the presence of current $I$, these are marked $B_m^{low}(T, I)$ curves. The red dashed line is the low field melting line with $I = 0$ mA, $B_m^{low}(I = 0$ mA$)$ see [22]. Black circle data represents a crossover above which current distribution in the sample becomes uniform. (b) Inset image is the MOI$_{SF}$ image taken at $T = 33$ K, $I = 30$ mA, and $B_a = 5$ G where the bright feature is current induced melting at low fields. In the main panel, solid blue circle is $\left(\frac{T_m(I)}{T_c}\right)_{B_m \to 0}$ vs $\frac{I}{I_c}$ plot where red (dashed) curve, and black (dotted) line are fit for $I > 50$ mA and $I < 50$ mA region, respectively. In all the plots, we have chosen the symbol size comparable to the size of the error bars.

It may be noted that the current densities $j$ used in our experiment ranges from 33 to 167 A.cm$^{-2}$, which is lower than the sample bulk critical current $J_c \sim 10^4$ A.cm$^{-2}$ [22]. Hence, the $I$ only drives the weakly pinned vortices in the sample, which are present in the regions of the sample with vortex solid (prior to melting here) and in the liquid regions. In the $B_a$-$T$ phase diagram of Fig. 3(a), we show with a dashed (red) curve the low-field melting line obtained with 0 mA, viz., $B_m^{low}(0$ mA$)$ line, shown in Fig. 4(c) of Ref. [22] [viz., locus of $(B_m, T_m)$ values at which melting occurs at the location marked as a red • in Fig. 1(a), for $I = 0$ mA]. Theories suggest [5,20] that in an ideal pinning free superconductor below $B_m^{low}(T)$ line, the dilute low-field liquid phase should extend all the way down to the lower critical field. However, Ref. [22] showed melting in this crystal is absent at very low $B_a$, and in fact a pinned glassy vortex state exists with finite $J_c$ prior to the vortex liquid phase near the $B_m^{low}(T)$ line (for $I = 0$ mA) (the relevant low field portion of phase diagram of Ref. 22, showing the presence of a low field glass below the liquid phase is reproduced in Supplementary Material Fig. 1). From Fig. 3(a) we see that the $B_m^{low}(T, 10$ mA$)$ and $B_m^{low}(T, 0$ mA$)$ lines deviate significantly above $T_m/T_c \sim$ 0.8. In fact, above a reduced temperature of 0.9 the $B_m^{low}(T, 10$ mA$)$ becomes very small (namely below our $B_a$ resolution of 1 G). For example, the self-field image in the inset of Fig. 3(b), shows the bright melted patch has already occurred for a $B_a$ as low as 5 G at 33 K i.e., $T/T_c \sim 0.87$ with $I = 30$ mA. Thus, with a current switched on, the vortex state begins to melt from very low fields. We see in Fig. 3(a), that the disordered low-field glassy phase, which was earlier present below the $B_m^{low}$ line for $I = 0$ mA [Ref. [22] and see Supplementary Material Fig. 1] is destabilized and disappears when a current is applied. With increasing $I$, we see the $B_m^{low}(T, I)$ line peels off from the $B_m^{low}(T, 0$ mA$)$ line at lower reduced $T_m$ values. We determine the $T_m$ values obtained at which $B_m^{low}(T, I) \to 0$ G, when $I$ is kept on. In Fig. 3(b) we plot $\left(\frac{T_m(I)}{T_c}\right)_{B_m \to 0}$ vs $\frac{I}{I_c}$ (the bulk $I_c \sim 10$ Amps as determined from the $J_c$ of the sample).

Consider $r$ is the local resistance of a region where locally $J_c$ is low due to which some vortices start moving under the influence of $I$. Due to this a local dissipation (Joule heating) occurs in these regions, whose value is: $I^2 r \Delta t \sim mC\Delta T$, where $\Delta t$ is the typical time current is kept on to capture a self-field image, $m$ is the mass of the crystal, $C$ is the specific heat and $\Delta T$ is the temperature difference due to Joule heating. Due to Joule heating melting would be observed at $T' < T_m$ as

$$T' = T_m(I=0) - \Delta T(I) = T_m(I=0) - \frac{I^2 r \Delta t}{mC} \qquad (1)$$

In Fig. 3(b) we see the data fits with Eq. 1 for $I \geq 50$ mA and the fit gives $\frac{r\Delta t I_c^2}{mCT_c} = 8.5 \times 10^3$. Using $\Delta t = 300$ s, $C \sim 18$ J.mole$^{-1}$K$^{-1}$[30], and $m \sim 1$ mg we estimate the local resistance (responsible for local dissipation) $r = 0.5$ mΩ (note bulk resistance of sample is ~5.6 mΩ). Figure 3(b) shows that the lowering of the melting temperature for $I \geq 50$ mA is due to Joule heating effects. However, below 50 mA, the suppression of the drive dependent melting $T_m$ in Fig. 3(b) is not explained by the Joule heating model. In this regime, we explore the suppression of drive dependent $T_m$ with $I$ using the dynamic vortex melting scenario [18,19]. At low drives, vortices driven over pinning centres experience a velocity fluctuation. In the frame of reference of the moving vortices, it will appear like a static system with vortices shaking about their mean location due to an effective shaking temperature $T_{sh}$ [18,19]. Due to $T_{sh}$, melting should occur at $T' = T_m(v=0) - T_{sh}(v) = T_m(v=0) - \frac{c_0}{v}$, where $c_0$ is a constant and $T_{sh}(v) \propto v^{-1}$ for low drives [18]. For our low $I \ll$ bulk $I_c$, we consider the vortices are in a thermally activated creep regime where vortices hop over pinning barriers of height $\Delta_0$. It was shown that in this creep regime [6] the vortex velocity scales with $F$ as [6,31], $v \propto e^{-k\left(\frac{F_c}{F}\right)^\gamma}$, where, $\gamma = \frac{d-2+2\eta}{2-\eta}$, $d$ is the dimension and $\eta$ is a scaling exponent. The quantity $k \sim \frac{\Delta_0}{k_B T}$ ($k_B$ is the Boltzmann constant) is treated as constant for this non-equilibrium driven situation, where the vortex velocity is primarily affected by the driving force $F \propto I$, which tilts the potential. Using the expression for $v$ in expression $T'$ and Taylor expanding exponential term, we get $T' = T_{m(v=0)} - c_1 e^{k\left(\frac{F_c}{F}\right)^\gamma} = (T_{m(v=0)} - c_1) - \chi[\left(\frac{F_c}{F}\right)^\gamma] +$ [higher order terms] where $\chi$ is a constant. Using $\gamma = -1$ and retaining only up to the first term in the above expression for low $I$ ($\ll I_c$), we get

$$\frac{T'}{T_c} = \frac{(T_{m(v=0)} - c_1)}{T_c} - \frac{\chi}{T_c} \cdot \left(\frac{I}{I_c}\right) = c_2 - c_3 \left(\frac{I}{I_c}\right) \qquad (2)$$

We see that below 50 mA the data in Fig. 3(b) fits well to Eq. 2, and we obtain the values of the constants $c_2 = 0.98$ and $c_3 = 37.5$. It may be noted that the higher-order terms in Eq. 2 are quadratic in $I$ and contribute only at high $I$, where we have already shown that Joule heating effects take over. We believe the effect of shaking temperature is responsible for destabilizing the low field glassy phase, which was present in the static phase diagram below the low field melting line [Ref. [22] and see Supplementary Material Fig. 1]. Thus, $I$ drives weakly pinned vortices, which either through the dynamic effect of shaking temperature or enhanced Joule heating (depending on the value of $I$) precipitates the early onset of vortex melting. As more depinned vortices are added to the system due to melting, the effect proliferates across the sample causing early onset of melting elsewhere in the sample.

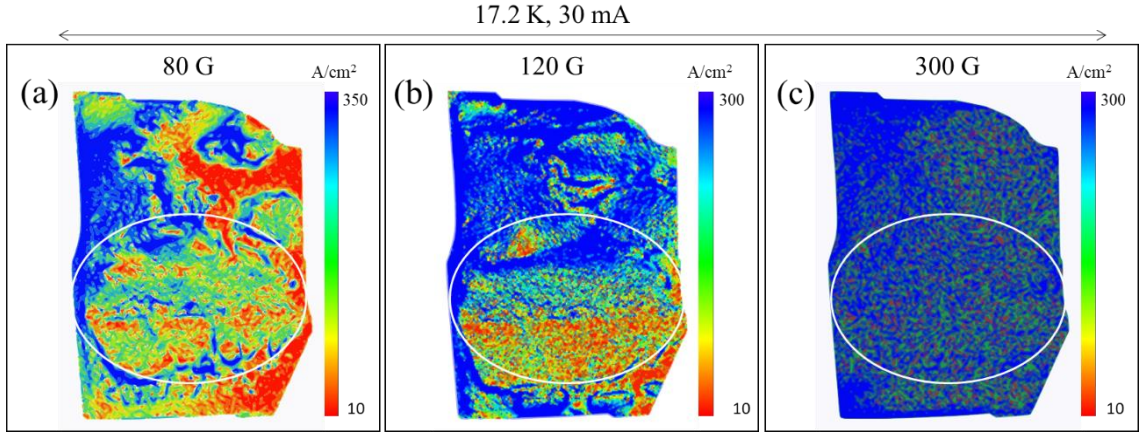

FIG. 4. Figures in (a)-(c) are maps of the current density distribution [$j(x, y)$] across the entire sample at $T = 17.2$ K for $I = 30$ mA and at different $B_a$. The white circled region shows the presence of significant current density inside the vortex liquid region.

Figure 4 shows the current density distribution across the sample as the vortex melting proceeds across the sample at 30 mA current. Inside the circled region of Fig. 4(a), we see an inhomogeneous current density distribution where one has a a liquid phase [see melting in this region has occurred at 80 G (@0 mA) in Fig. 3(a)]. We believe this inhomogeneity is due to a spatially inhomogeneous $T_{sh}$, which in turn reflects the non-uniformity of the underlying pinning landscape in the sample. With higher driving force (higher $B_a$ for same $I = 30$ mA) as $T_{sh} \propto \frac{1}{v}$, the non-uniformity of $T_{sh}$ weakens as seen and $j(x, y)$ becomes less inhomogeneous [cf. 120 G data (Fig. 4(b)]. At large drives, vortices are weakly affected by the underlying pinning landscape, hence $T_{sh} \to 0$ at large drives. Deep in the solid phase at 300 G where the Lorentz force drive is high the role of $T_{sh}$ ceases and $j(x, y)$ is homogeneous [Fig. 4(c)]. This inhomogeneous to homogenous $j$ distribution crossover we believe verifies the inverse $T_{sh}$ dependence on vortex velocity.

In conclusion, we show drive affects phase transitions in the vortex state and the role of shaking temperature in modifying some of the quasi-equilibrium phases and transformation boundaries. Some of these studies pave the way for more detailed theoretical and experimental investigations into the study of non-equilibrium condensed matter systems.

**Acknowledgement:** S.S.B. would like to acknowledge the funding support from IITK(IN) and DST-TSDP(IN), Govt. of India.

# Supplementary Material

# Imaging the effect of drive on the low-field vortex melting phenomenon in $Ba_{0.6}K_{0.4}Fe_2As_2$ single crystal

Ankit Kumar[1], Amit Jash[1], Tsuyoshi Tamegai[2], S. S. Banerjee[1,*]

[1]Department of Physics, Indian Institute of Technology, Kanpur-208016, India

[2]Department of Applied Physics, The University of Tokyo, Hongo, Bunkyo-ku, Tokyo 113-8656, Japan

*Email: satyajit@iitk.ac.in

## $B_m(T)$ phase diagram showing low-field glassy and liquid vortex phases

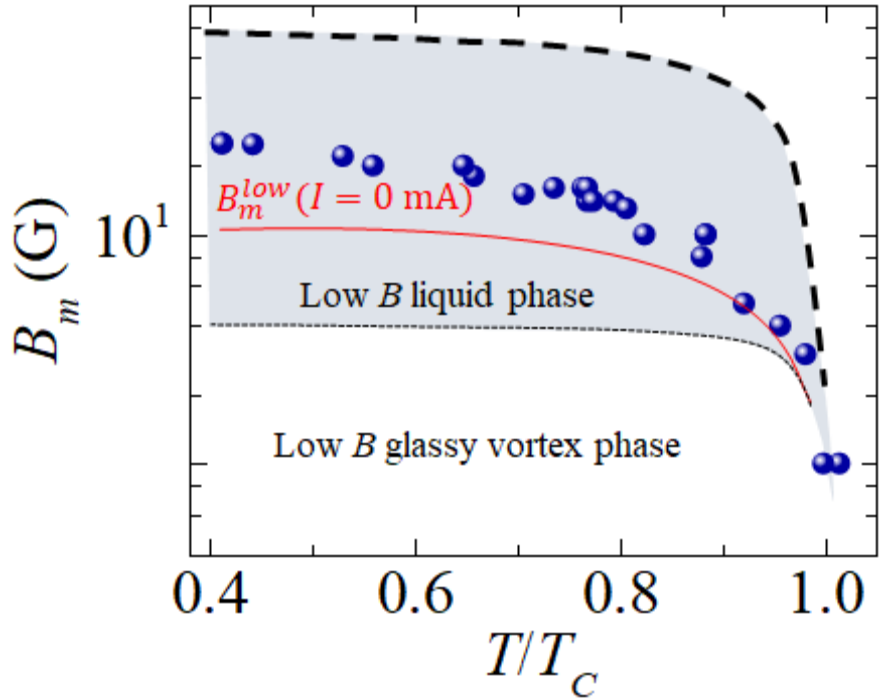

FIG. 1. $B_m(T)$ phase diagram (blue color data) is plotted for the sample region marked as • in Fig. 1(a) [1]. The red color solid line is a fit to the low-field melting line ($B_m$) [2]. A shaded region has been shown between the two-dashed line to distinguish between a liquid-like phase from a low-field vortex glass and a soft vortex solid phase (not shown) at higher fields. Low field glassy vortex phase occurs due to weak rigidity of the vortex state at low fields in the presence of the naturally occurring strong extended pinning centers in the sample.

## DMO images showing the evolution of vortex melting

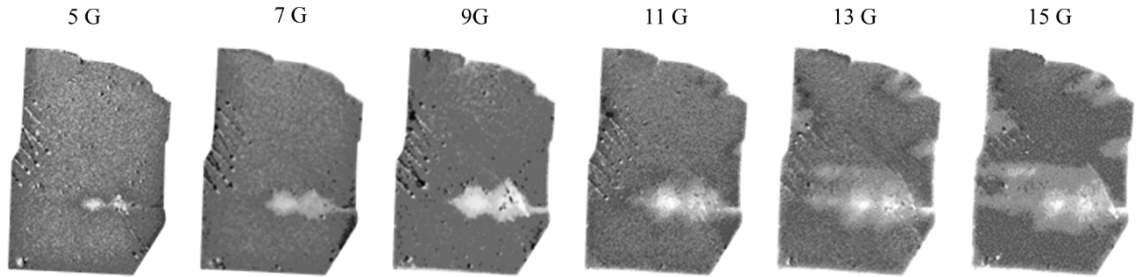

FIG. 2. Shows the differential magneto-optical (DMO) images taken at 32 K, $I = 0$ mA and at different $B_a$. The grey regions have $\delta B_z = \delta B_a = 1$ G, wherein the density of vortices follows the changes in the external magnetic field. Notice the bright regions in the DMO images occur in the same sample location where the brightening was seen in Fig. 1(a). These bright regions are not symmetric patches but possess a directionality in their shape. Over the bright regions in DMO images, the $\delta B_z$ is larger than 1 G [1], and hence in these regions the local vortex density has changed more in comparison to other neighbouring grey regions. In all the images, left edge show a zigzag pattern. This zigzag pattern is due to well-known Bloch walls seen on the magneto-optical film [3], which is placed on top of the sample for magneto-optical imaging.